\documentclass{WileyMSP-template}
\usepackage{textcomp}
\usepackage{float}
\usepackage[numbers,sort&compress]{natbib}  
\usepackage{amsmath}
\usepackage{xcolor} 
\usepackage[normalem]{ulem}

\begin{document}

\justifying


\title{Thouless Pumping of \textcolor{black}{One-Dimensional} Dark Solitons and Domain Walls}

\maketitle


\author{Fangwei Ye$^{1,2,*}$}
\author{Aidar H. Ryazhapov$^{3,4}$}
\author{Yaroslav V. Kartashov$^{3}$}
\author{Vladimir V. Konotop$^{5}$}

\begin{affiliations}
$^{1}$School of Physics, Chengdu University of Technology, Chengdu 610059, China
\end{affiliations}

\begin{affiliations}
$^{2}$School of Physics and Astronomy, Shanghai Jiao Tong University, Shanghai 200240, China
\end{affiliations}

\begin{affiliations}
$^{3}$Institute of Spectroscopy, Russian
Academy of Sciences, 108840 Troitsk, Moscow, Russia
\end{affiliations}

\begin{affiliations}
$^{4}$Faculty of Physics,
Higher School of Economics, 105066 Moscow, Russia
\end{affiliations}

\begin{affiliations}
$^{5}$Departamento de Física and Centro de
Física Teórica e Computacional, Faculdade de Ciências, Universidade de Lisboa, Campo Grande, Edifício C8, Lisboa
\end{affiliations}

{{*}Corresponding author: \noindent Fangwei Ye,  {fangweiye@sjtu.edu.cn} }

\
\

\textbf{Abstract}:
We predict that quantized transport of dark solitons and domain walls can be observed in mutually sliding optical lattices in both focusing and defocusing nonlinear materials. Two different classes of underlying periodic backgrounds are considered here -- unstaggered and staggered -- which are accurately represented, respectively, by the in-phase and out-of-phase trains of Wannier functions. The nonlinear excitations formed on top of these backgrounds -- dark solitons and domain walls -- are sustained by defocusing or focusing Kerr nonlinearities of the optical medium. Despite non-zero asymptotics, dark solitons and domain walls can still undergo quantized topological transport in dynamically changing refractive index landscape when they and background on which they are constructed are dynamically stable. The extended intervals of propagation constants (and, correspondingly, of the background amplitude) and lattice depths, where stable quantized transport occurs, are identified for both dark solitons and domain walls. Outside these intervals, the quantized transport regime is found to be unstable. Our results extend the concept of nonlinear topological transport from bright to dark soliton states.

\section{Introduction}

Adiabatic quantized transport of solitons has recently attracted substantial interest, particularly in the context of Thouless pumping~\cite{Thouless1983} of optical beams and matter waves. It has been realized~\cite{FuPRL2022a,JurgensenPRL2022} that this fundamental phenomenon can occur in nonlinear medium, where it is still determined by the topological invariants of excited bands of the underlying linear system, even in strongly nonlinear regime, where interband transitions are possible. In discrete settings, this phenomenon was demonstrated experimentally in arrays of optical waveguides~\cite{JurgensenNature2021}, and analyzed theoretically in discrete lattice models~\cite{JurgensenPRL2022, Mostaan2022, Ravets2025}. Its realization in fully continuous systems was predicted and described in Ref.~\cite{FuPRL2022a}. In subsequent studies, quantized transport of a rich diversity of solitons have been reported. Specifically, topological transport has been shown to occur for vector solitons in two-component Bose-Einstein condensates~\cite{Lyu2024,Cao2025}, solitons in spin-orbit coupled Bose-Einstein condensates~\cite{Kartashov2026}, multifrequency solitons in quadratic optical materials~\cite{Kartashov2025}, and solitons in nonlocal media~\cite{Ye2025APL}, see also review \cite{Szameit2024}.
 
Generally, dynamics of Thouless pumping of solitons is characterized by two independent factors. The first one is the conventional (i.e., linear) topology of the underlying periodic medium in the extended coordinate-time space (here, time is understood in a generalized sense as an evolution coordinate). The second mechanism is the interband transfer of energy (or particles), governed by the nonlinear coupling between the bands of the linear lattice. Due to such coupling, the transport of solitons can acquire very unusual features. For instance, it may become fractional, i.e., exhibiting a shift over only a certain fraction of the spatial period during one pumping cycle \cite{JurgensenNature2021, FuPRL2022a, Tao2025}. Furthermore, Thouless pumping of solitons has been demonstrated in lattices containing impurities~\cite{Cao2024}, as well as in two-dimensional lattices subject to adiabatic parameter modulation~\cite{FuPRL2022b, Ye2026}.  

Up to now nonlinear Thouless pumping was considered exclusively for localized self-sustained states, i.e. for bright solitons. The fundamental question, whether this phenomenon is possible for dark solitons and nonlinear states with nonzero asymptotics remains unanswered. Very recently it was proposed \cite{TaoDark} that quantized transport is principally possible for dark solitons in lattices, a class of models that plays a significant role in nonlinear optics~\cite{Lederer2008, Kartashov2011} and in theory of Bose-Einstein condensates \cite{Brazhnyi2004}. This proposal relies on the observation that the background associated with a dark soliton can be interpreted as an array of Wannier states and can therefore be pumped in dynamically changing potential. Although the conceptual framework was outlined in~\cite{TaoDark}, it was based solely on the possibility to find "instantaneous" self-sustained dark states at different distances in dynamically varying potential, without considering their actual dynamical evolution. Thus, underlying mechanisms and conditions for occurrence of {\em stable} quantized transport of dark solitons remain largely unexplored.

Indeed, the first fundamental problem that arises in theory of nonlinear Thouless pumping is connected with stability of excitation in dynamical potentials and it is particularly important for dark solitons, since they (and periodic background on which they reside) may exhibit dynamical instabilities~\cite{Bronski2001, Konotop2002, Brazhnyi2004, Lederer2008, Kartashov2011}, necessitating an analysis in the real-time framework (i.e., solving a time-dependent problem). In contrast, a purely parametric (time-independent) analysis does not allow for drawing any conclusions regarding soliton and background stability. Second, the use of Wannier functions, which in the corresponding limit leads to a discrete nonlinear Schrödinger equation~\cite{AKKS}, naturally highlights the analogy between dark solitons in deep continuous lattices and in discrete lattices. Meanwhile, it is known that the latter can support at least four, rather than only one, types of (static) dark solitons, characterized by either on-site or inter-site position of the center and by either staggered or unstaggered backgrounds \cite{Kivshar1994, Johansson1999, Konotop1999, Kevrekidis2003, Susanto2005}. Similar four types of solitons exist also in continuous models (see studies in matter-wave \cite{Louis2004} and optical \cite{Morandotti2001, Smirnov2006, Dong2010, Kanshu2012} systems) and therefore can potentially undergo pumping. Finally, one can explore an “intermediate” type of solution — domain walls, which approach different asymptotic limits at infinities: zero on one side and a nonlinear periodic wave (i.e., a nonzero background) on the other. Pumping of such states has never been addressed so far and its study constitutes one of our main goals.

In this work, we show that both dark solitons and domain walls constructed on staggered and unstaggered backgrounds can exhibit topological transport in dynamically changing continuous optical potential created by two mutually sliding sublattices. Stable pumping is possible both in defocusing (for solitons with unstaggered background) and in focusing (for solitons with staggered background) materials. The domains of stable pumping in terms of propagation constants and lattice depths are identified for all types of solutions. Our results hint on the fact that fundamental phenomenon of topological transport can be observed for much wider class of nonlinear waves as it was believed before, suggesting particularly interesting extensions to multidimensional settings and solutions with nontrivial structure of phase front.

\section{The model and nonlinear background}

While our results are applicable to a wide range of optical and BEC settings, to be specific, in this work we explore the model
\begin{align}
\label{NLS}
    i\frac{\partial\Psi}{\partial  z}=-\frac{1}{2}\frac{\partial^2\Psi}{\partial x^2}-\mathcal{R}(x,z)\Psi+\chi |\Psi|^2\Psi
\end{align}
where the pumping for light field with dimensionless amplitude $\Psi$ is implemented through the optical potential consisting of two mutually sliding sublattices
\begin{align}
\label{potential}
    \mathcal{R}(x,z)=p_1\cos^2(\pi x/d_1)+p_2\cos^2[\pi (x-\alpha z)/d_2]
\end{align}
with the amplitudes $p_{1,2}$ and commensurable periods $d_{1,2}$ with $X$ being the common period, and sliding angle $\alpha\ll 1$. In this model, $x$ is the normalized transverse coordinate, $z$ is the normalized propagation distance, the optical potential $\mathcal{R}$ is created by the transverse and longitudinal shallow modulation of the refractive index of the transparent optical medium, $\chi$ is the nonlinearity coefficient ($\chi<0$ in focusing medium, and $\chi>0$ in defocusing medium). We use here standard normalizations that can be found in \cite{Kartashov2011}. 

\begin{figure}[t]
\begin{center}
			\includegraphics[width=0.5\columnwidth]{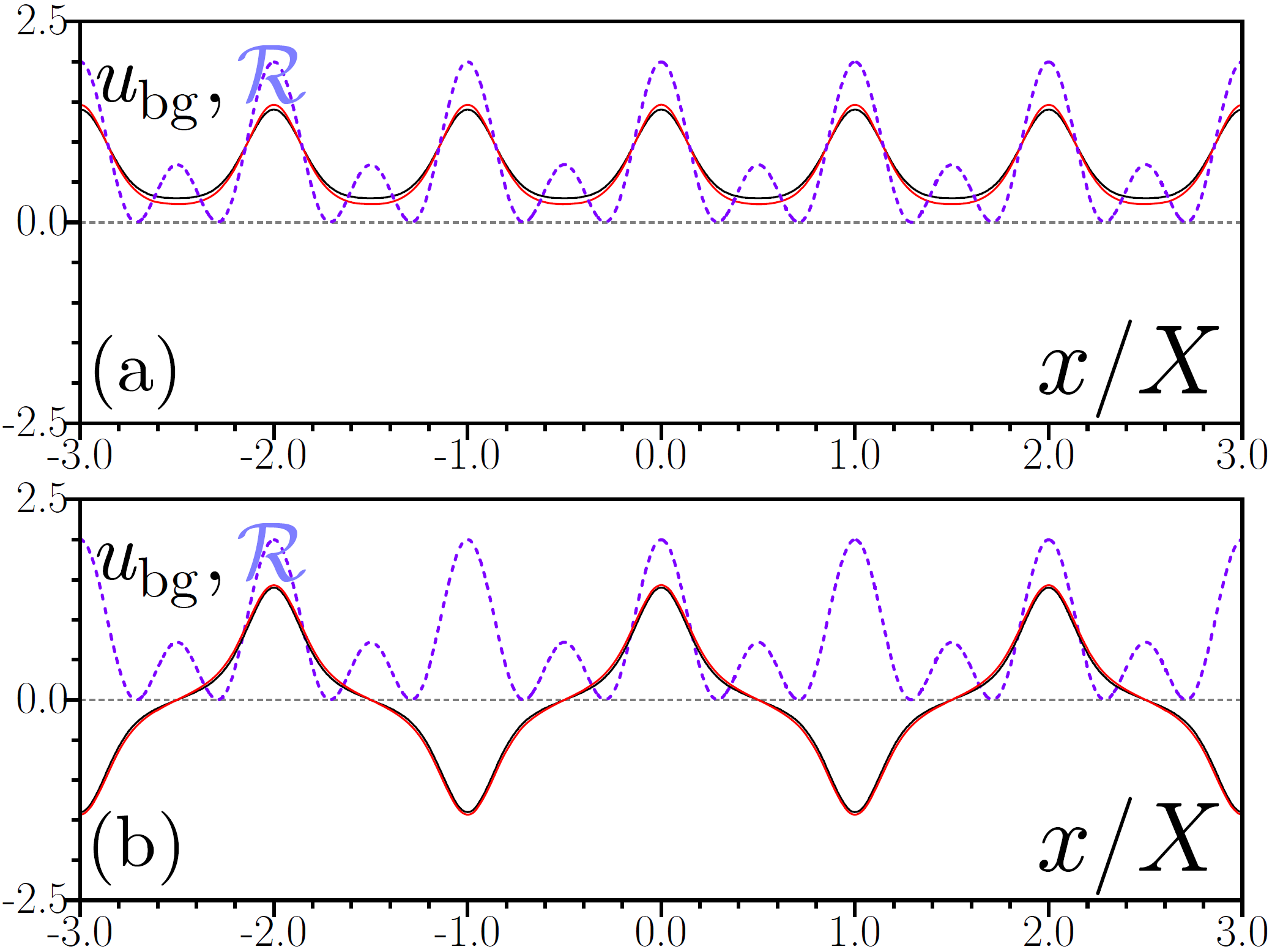}
\end{center}
\caption{Modulated unstaggered background in a defocusing medium at $b = 9.0$ (belonging to the first finite gap in spectrum of $\mathcal{R}$) (a) and staggered background in a focusing medium at $b = 11.5$ (belonging to the semi-infinite gap in spectrum of $\mathcal{R}$). Black lines show exact solutions, while red lines show the reconstructed background from Wannier functions from the first band. Dashed violet line schematically shows the profile of the lattice $\mathcal{R}$ at $z = 0$. In all cases $p_{1,2} = 8$, $d_1 = 1$, $d_2 = 2$.}
\label{fig1}
\end{figure}

Dark lattice solitons, whose pumping will be considered in this work, are characterized by non-vanishing asymptotic behavior at the infinities $x\to\pm\infty$, i.e., far from the soliton center the solutions tend a nonzero periodic background. At the input plane $z=0$, the profile of the background can be obtained from Eq.~(\ref{NLS}) in the form $\Psi_{\rm bg}(x,z)=u_\textrm{bg}(x)e^{ibz}$, where $u_\textrm{bg}(x)$ is real-valued function and $b$ is the propagation constant. This background can be either unstaggered [i.e. $u_\textrm{bg}(x)$ has the same phase at neighboring main lattice maxima] or staggered [the phase of $u_\textrm{bg}(x)$ changes by $\pi$ between neighboring main lattice maxima]. A typical example of the unstaggered background at $z=0$ in defocusing nonlinear medium is presented in Fig.~\ref{fig1}(a) (black line) where the propagation constant $b$ was chosen within first finite gap of the linear lattice spectrum. An example of staggered background in focusing nonlinear medium is presented in Fig.~\ref{fig1}(b) (black line) where the propagation constant $b$ is chosen in semi-infinite gap of the linear lattice. The correspondence between the background type and the sign of the nonlinearity is essential: only a staggered background can be stable in a focusing medium, whereas only an unstaggered background can be stable in a defocusing medium. Consequently, in what follows we restrict our consideration to dark solitons and domain walls constructed on unstaggered (staggered) backgrounds in defocusing (focusing) media, respectively.

The asymptotic behavior of a dark soliton at infinity is particularly significant in the presence of the $z$-dependent potentials given by (\ref{potential}), and is conceptually distinct from its analogue in the parametric setting (i.e., when the evolution coordinate is treated as a mere parameter). Specifically, once the explicit $z$-dependence is taken into account, the background field on which a dark soliton can be constructed must necessarily become complex-valued. To demonstrate this, we consider a general \textit{evolving} background solution of Eqs.~(\ref{NLS}), (\ref{potential}) of the form $\Psi(x,z) = w(x,z)e^{i\theta(x,z)}$, where $w$ and $\theta$ are real-valued functions describing, respectively, the amplitude and phase of the solution:
\begin{align}
 w(x+X,z)=w(x,z), \qquad \theta(x+X,z)= \theta(x,z)+\theta_0, \qquad  w(x,0)=u_{\rm bg}(x)  
\end{align}
The phase shift $\theta_0\in[0,2\pi)$ determines two types of the nonlinear background considered in this work: the unstaggered background corresponding to $\theta_0=0$ and staggered background corresponding to $\theta_0=\pi$. 

Since our focus is on the adiabatic limit $(\alpha\ll 1)$, we introduce the slow variable $\zeta = \alpha z$ and a solution on the evolving background in the form of the expansions
\begin{align}
    w=& w_0(x,\zeta)+\alpha^2w_2(x,\zeta)+\cdots
    \\
    \theta =&\frac{1}{\alpha}\int b(\zeta) d\zeta   +\alpha\vartheta(x,\zeta)+\cdots
\end{align}
Separating real and imaginary parts, in the zero $(\alpha^0)$ and first $(\alpha^1)$ orders we obtain
\begin{align}
\label{background}
    -\frac{1}{2}\frac{\partial^2w_0}{\partial x^2}+[b(\zeta)-\mathcal{R}(x,\zeta)]w_0+\chi w_0^3=0
    \\
    \label{phase}
    w_0\frac{\partial^2\vartheta}{\partial x^2}
    +2\frac{\partial w_0}{\partial x}\frac{\partial\vartheta}{\partial x}+2\frac{\partial w_0}{\partial \zeta}=0
\end{align}
The dependence of $w_0$ on $\zeta$ inhibits solutions with $\vartheta$ independent of $x$, which means that for any nonzero $\alpha>0$ the background of evolving solution of Eq.~(\ref{NLS}) must be complex. In particular, from (\ref{phase}) one can obtain
\begin{align}
\label{theta_x}
    w_0^2\frac{\partial\vartheta}{\partial x}=-   \int_{x_0}^x \frac{\partial w_0^2}{\partial \zeta}  dx 
\end{align}
where $x_0$, which generally speaking, is a function of $\zeta$, must be determined depending on the choice of $\theta_0$. To this end we note, that for unstaggered background, that is strictly positive, $w(x,z)>0$ [Fig.~\ref{fig1}(a)], periodicity implies that there exists at east one point $x_1(\zeta)$ on each spatial period in which $\vartheta_x=0$. In that case one can set $x_0=x_1(\zeta)$ in (\ref{theta_x}). In the case of a staggered background [Fig.~\ref{fig1} (b)], for each $\zeta$, the amplitude $w_0 (x,\zeta)$ vanishes within each period of the background. Denoting this point by $x_2(\zeta)$, we have $w_0(x_2(\zeta),\zeta)=0$. Hence, one can set $x_0=x_2(\zeta)$. It should also be mentioned that the power of the background per period  is conserved, i.e.,  $\int_{x_0}^{x_0+X} w_0^2 dx$ is $\zeta-$independent, and $\partial w_0^2/\partial \zeta$ is $X-$periodic, what guarantees $X-$periodicity of the derivative $\vartheta_x$. This analysis directly shows that to adequately describe dynamics of Thouless pumping of dark solitons, one has to use time-dependent framework. In particular, one must model evolution of solitons along $z$ by Eq.~(\ref{NLS}), that accounts for the appearance of nontrivial phase in the course of evolution.

For the purposes of the present paper, we expand the background over the complete set of the Wannier functions $w_{\nu n}(x,z)$ (see \cite{Kohn1959}) associated with the potential (\ref{potential}):
\begin{align}
\label{expan_Wan}
	w(x,z)=\sum_{\nu=1}^\infty\sum_{n=-\infty}^\infty a_{\nu n}(z) w_{\nu n}(x,z)
\end{align}
Here, $\nu$ is the band index in the band-gap spectrum of the lattice $\mathcal{R}$ (chosen such that $\nu=1$ corresponds to the uppermost band), whereas $n$ labels the lattice cell (with the cell $n=0$ centered at $x=0$). For sufficiently large depths of the sublattices, $p_{1,2}$, like those used in this paper, the above expansion remains quantitatively highly accurate when the Wannier functions are restricted to those associated with the top band, i.e., to $\nu=1$ only, since absolute values of $a_{\nu n}$ rapidly decay with $\nu$. Thus, for the unstaggered background shown in panel (a) we computed $a_{1n} \approx 1.1247$, $a_{2n} \approx 6.92\cdot 10^{-2}$, and $a_{3n} = 0$, while for the staggered background shown in panel (b) one has $a_{1n} =(-1)^n 1.0434$, $a_{2n} =(-1)^n 3.73 \cdot 10^{-3}$, and $a_{3n} =0$. In Fig.~\ref{fig1}(a),(b), the red line represents the reconstruction of the exact background profile $u_\textrm{bg}(x)$ at $z=0$ obtained using only $\nu=1$ Wannier functions. This reconstruction closely approximates the exact state and thus permits, to leading order, an interpretation of the background as a train of independent Wannier functions that are effectively pumped when the sublattices constituting $\mathcal{R}$ slide relative to each other.

\section{Dark solitons and domain walls} 

\begin{figure}[t]
\begin{center}
			\includegraphics[width=0.5\columnwidth]{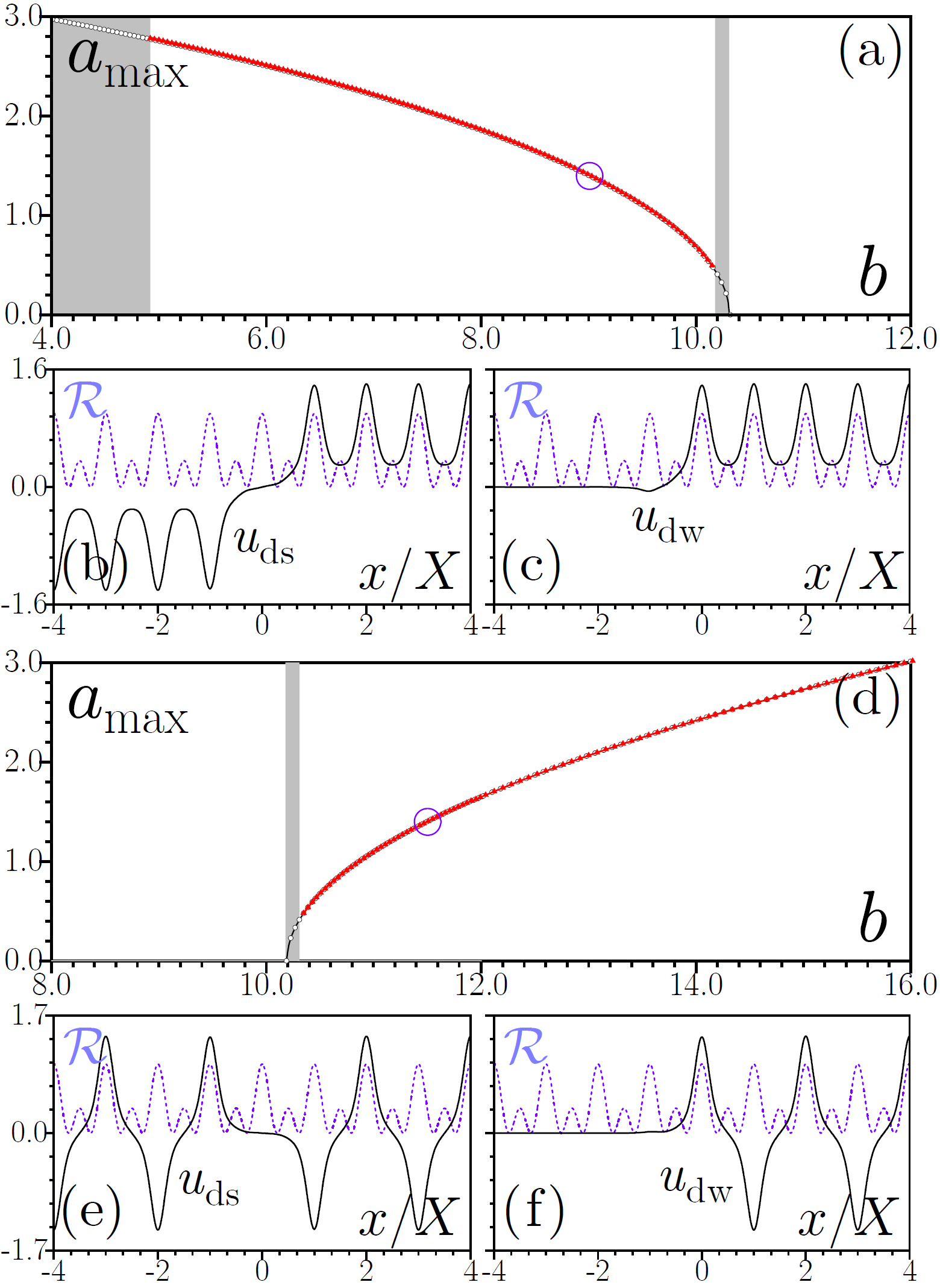}
\end{center}
\caption{Amplitude of dark soliton (open circles) and domain wall (red triangles) vs propagation constant $b$ in defocusing (a) and focusing (d) media, obtained for $z=0$. Gray areas correspond to allowed bands. Inside the forbidden gap, the families of dark solitons and domain walls are practically non-distinguishable on the scale of the figures. Profiles of unstaggered dark soliton (b) and domain wall (c) with $b = 9$ in a defocusing medium. Profiles of staggered dark soliton (e), and domain wall (f) with $b = 11.5$ in a focusing medium. Dashed violet line schematically shows the profile of the lattice $\mathcal{R}$ at $z = 0$. In all cases $p_{1,2} = 8$, $d_1 = 1$, $d_2 = 2$.}
\label{fig2}
\end{figure}

Next we proceed to properties of dark solitons and domain walls. Single dark soliton represents a solution of Eq. (\ref{NLS}) with optical potential (\ref{potential}) of the form $\Psi(x,z)=e^{i\int b(z)dz}\psi_{\rm ds}(x,z)$ satisfying the boundary conditions
\begin{align}
\label{soliton}
    \lim_{x\to\pm\infty} \psi_{\rm ds}(x,z)=\pm w(x,z)
\end{align}
As shown above, upon pumping process in dynamical potential landscape this solution (that asymptotically tend to modulated background) should acquire inhomogeneous in $x$ phase. This is purely dynamical, i.e., non-adiabatic, effect contrasting with assumptions of \cite{TaoDark}. The fact that at certain instance in $z$ such phase can be set to zero, allows us to define the initial conditions in the form of purely real "on-site" dark soliton (which at $z=0$ is centered at $x=0$):
\begin{align}
\label{soliton_0}
    \psi_{\rm ds}(x,0)=u_{\rm ds}(x), \quad   \lim_{x\to\pm\infty} u_{\rm ds}(x)=\pm w(x,0), \quad {\rm Im}\,u_{\rm ds}(x)=0
\end{align}
In defocusing nonlinear medium such dark initial excitations constitute a family, which bifurcates from the upper edge of the allowed band with index $\nu=1$. This family can be described by the dependence of maximal amplitude of solution $a_\textrm{max}=\textrm{max}|u_\textrm{ds}|$ (that in fact coincides with amplitude of the unstaggered background) on propagation constant $b$ presented in Fig.~\ref{fig2}(a) (see curve with open circles). An example of "on-site" dark soliton is presented in Fig. \ref{fig2}(b). we note that the family of dark solitons exist within $\nu=1$ band (right gray region), then it enters the first gap (white region), and after crossing it continues into the $\nu=2$ band (left gray region). Thus, such solitons exist both in gaps and allowed bands. Notice that central notch in dark soliton becomes broader with increase of propagation constant $b$.

The second type of solutions whose pumping is explored in this paper is represented by domain walls having different asymptotics
\begin{align}
\label{wall}
    \lim_{x\to\mp \infty} \psi_{\rm dw}(x,z)=0, \quad \lim_{x\to\pm\infty} \psi_{\rm dw}(x,z)= w(x,z)
\end{align}
Just as in the case of dark solitons, the initial conditions representing domain walls can be taken as purely real solutions at $z=0$, with domain wall initially located in the point $x=0$. In this case, one has
\begin{align}
\label{wall_0}
    \psi_{\rm dw}(x,0)=u_{\rm dw}(x), \quad   \lim_{x\to\mp \infty} u_{\rm dw}(x)=0, \quad \lim_{x\to\pm\infty} u_{\rm dw}(x)= w(x,0), \qquad {\rm Im}\,u_{\rm dw}(x)=0
\end{align}
An example of input profile of the domain wall at $z=0$ in defocusing medium is shown in Fig.~\ref{fig2}(c). Because such domain wall has zero asymptotics at $x\to -\infty$, it can exist only within forbidden gap. The corresponding family represented by $a_\textrm{max}=\textrm{max}|u_\textrm{dw}|$ vs $b$ dependence is shown in Fig.~\ref{fig2}(a) with red triangles. Notice that while corresponding maximal amplitude is very close to that of dark soliton, it does not coincide with it exactly (these amplitudes differ near the top edge of the gap). Near both edges of the gap the domain wall acquires long oscillating tail inside $x<0$ region. The width of transition region in domain wall also increases with increase of propagation constant $b$.

Similar types of initial states in the form of "on-site" dark solitons and domain walls can be found also in focusing nonlinear medium, but to be potentially stable in such media, such solutions have to be constructed on staggered background. The family of staggered dark solitons bifurcates from the bottom edge of the $\nu=1$ band, crosses this band, and continues into semi-infinite gap, as shown with open circles in Fig.~\ref{fig2}(d). Representative profile of dark soliton with staggered background is shown in Fig.~\ref{fig2}(e). Typical staggered domain wall in focusing medium is depicted in Fig.~\ref{fig2}(f). The family of such domain walls exists within semi-infinite gap, as shown with red triangles in Fig.~\ref{fig2}(d). The formulas (\ref{soliton})--(\ref{wall_0}) remain valid in focusing medium, but $w(x,0)$ in them represents the staggered background [like the one shown in Fig.~\ref{fig1}(b)]. It should be mentioned that here we consider only "on-site" dark solitons, since they can exhibit stable pumping, while their "inter-site" variant, which can also be constructed, is usually unstable.

\section{Topological pumping and instabilities}
\label{sec:pumping}

Now we turn to transport of dark solitons and domain walls in the varying potential $\mathcal{R}(x,z)$. The relation between transport of the particular type of nonlinear state and the topology of the lattice $\mathcal{R}(x,z)$ is provided by the expansion of this state over the basis of the Wannier functions (\ref{expan_Wan}). For all states considered in this work, which bifurcate either from the upper or lower edge of the $\nu=1$ band, the Wannier functions associated with the upper $\nu=1$ band provide the dominant contribution to the corresponding modal expansion (as illustrated for both unstaggered and staggered nonlinear backgrounds in Fig.~\ref{fig1}). Consequently, the transport properties of these states are governed primarily by the topological characteristics of this band. 

For the chosen configuration with equal sublattice periods $d_{1} = d_{2}$, the Chern number $C_\nu$ of the upper band, calculated in the "coordinate–time" space~\cite{Thouless1983}, is found to be $C_{1} = +1$ for all considered sublattice depths (for the lower bands of this lattice the Chern numbers alternate, specifically $C_{2} = -1$ and $C_{3} = +1$). Therefore, provided that the initial nonlinear state is dynamically stable, one anticipates its transverse displacement by the distance $C_{1}X = X$ over a single pumping cycle of duration $Z = X/\alpha$.

\begin{figure}[t]
\begin{center}
			\includegraphics[width=0.6\columnwidth]{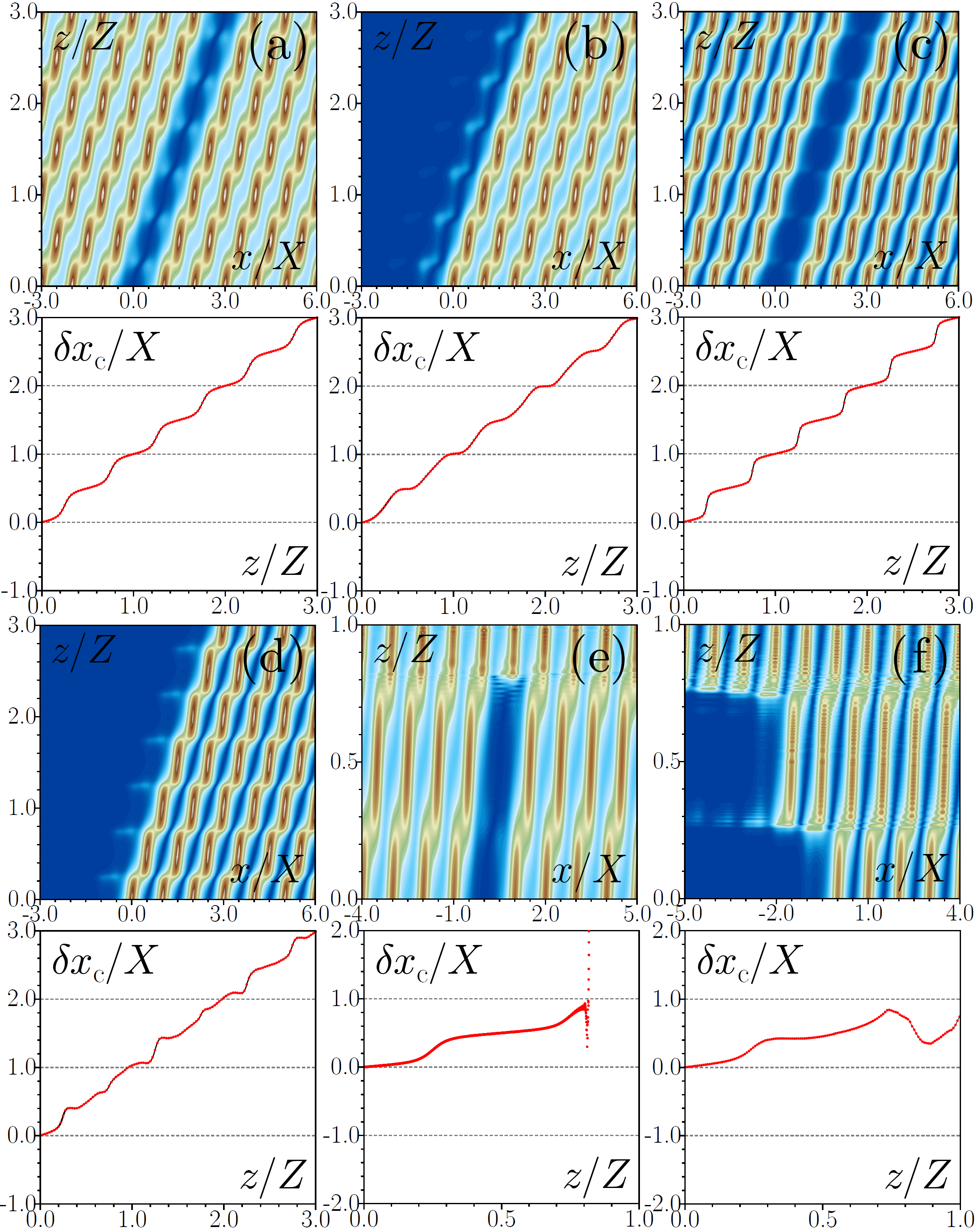}
\end{center}
\caption{Stable Thouless pumping of unstaggered dark soliton with $b = 8.0$ (a) and domain wall with $b = 9.0$ (b) in defocusing medium, and of staggered dark soliton (c) and domain wall (d) with $b = 11.5$ in focusing medium. Unstable pumping leading to decay of (e) \textcolor{black}{unstaggered dark soliton with $b = 9.3$} in defocusing medium and (f) \textcolor{black}{staggered domain wall with $b = 10.9$} in focusing medium. Top row shows dynamics of $|\Psi|$ in the $(x,z)$ plane, bottom row shows displacement of center of mass $\delta x_c$ vs distance $z$. Three pumping cycles are shown \textcolor{black}{in panels (a)-(d) with stable dynamics, and only one cycle is shown in panels (e),(f) showing unstable pumping}. Sliding velocity of sublattices is $\alpha = 0.001$.} 
\label{fig3}
\end{figure}

To illustrate this we model the direct evolution of the initial states (\ref{soliton})--(\ref{wall_0}) by the nonlinear Schrödinger equation (\ref{NLS}) using split-step fast-Fourier transform method. Our central result is that it is indeed possible to find parameter domains, where stable quantized transport of the on-site dark solitons and domain walls is possible. The examples with stable transport dynamics of such states in defocusing medium are presented in top panels of Fig. \ref{fig3}(a) and \ref{fig3}(b), respectively. In these figures we show the evolution of respective states over three pumping cycles, $z=3Z$. Notice that within stable transport region the solution undergoes exceptionally regular evolution, with its modulus distribution $|\Psi|$ nearly exactly reproduced after each pumping cycle (except for the quantized transverse shift of the entire distribution), without any obvious radiation. Such transport is remarkably robust and can persist over hundreds of $Z$ cycles. The bottom panels in Fig. \ref{fig3}(a) and \ref{fig3}(b) show the shift of the center of solution $\delta x_c=x_c(z)-x_c(0)$, where for dark soliton the instantaneous center position was defined as
\begin{align}
\label{shiftdark}
x_c=\frac{1}{N}\int [|w(x,z)|^2-|\psi_{\rm ds} (x,z)|^2]^fxdx, 
\end{align}
where $N=\int [|w(x,z)|^2-|\psi_{\rm ds} (x,z)|^2]^fdx$, while in the case of domain wall we have constructed the localized, but very broad input with two domain walls at opposite ends separated by very large distance $100X$ (to guarantee that such domain walls do not affect each other), in which case we defined position of center as
\begin{align}
\label{shiftwall}
x_c=\frac{1}{N}\int |\psi_{\rm dw} (x,z)|^{2f}xdx, 
\end{align}
where $N=\int |\psi_{\rm dw} (x,z)|^{2f}dx$, i.e. in this last case we do not subtract the modulated background from the evolving state. \textcolor{black}{In the mean-field framework, the definition of the center of mass of a dark soliton given by Eq. (\ref{shiftdark}) presupposes the dynamical stability of the background field $w(x,z)$. Because soliton instabilities are, in general, distinct from background instabilities, i.e., a soliton may exhibit instability even when the background remains stable, $x_c$ in Eq. (\ref{shiftwall}) remains well defined for both stable and unstable dark-soliton states. For domain walls, the quantity $x_c$ introduced in Eq. (\ref{shiftwall}) is similarly applicable to both stable and unstable solutions. Nevertheless, in both settings, $x_c$ yields physically informative diagnostics primarily under stable propagation; in unstable regimes it should instead be regarded as one of several integral measures that can be used to quantify the ensuing evolution.}

We further observe that, during the transport of dark states, small-amplitude traveling excitations develop atop the background across the full spatial extent of the system. These excitations originate from longitudinal variations of the lattice potential $\mathcal{R}$ and can yield a non-negligible contribution to the computed center position $x_c$, since the definition of the center assigns enhanced weight to contributions from large $x$. To reduce the impact of these linear excitations, we set $f=3$ in Eqs. (\ref{shiftdark}) and (\ref{shiftwall}).

In bottom panels of Fig.~\ref{fig3}(a) and \ref{fig3}(b) one clearly observes a quantized shift $\delta x_c=3X$ of the center of respective nonlinear states over three pumping cycles $z=3Z$. \textcolor{black}{It should be mentioned that upon stable pumping of the domain wall shown in Fig. \ref{fig3}(b) one can observe the appearance of a weak modulated stripe near the edge of the moving domain wall. This is a consequence of dynamical variation of the lattice and the fact that domain wall in defocusing medium [see example in Fig. \ref{fig2}(c)] has staggered decaying tail on the one side, whose amplitude may periodically increase and decrease upon periodic variations of $\mathcal{R}(x,z)$ as seen in Fig. \ref{fig3}(b).}

Stable topological transport is also possible in focusing medium, but now for states with staggered background. The example of such transport is presented in the top panel of Fig.~\ref{fig3}(c), while dynamics of transport of staggered domain wall is shown in the top panel of Fig.~\ref{fig3}(d). In both cases the shift of center of the pattern after each cycle is also clearly quantized, as one can see from respective bottom panels. This clearly demonstrates that the topological nature of transport for both dark solitons and domain walls is connected with dynamics of modulated background, and persists for different specific local features of the density profiles. \textcolor{black}{Notice that in focusing medium the tail of the domain wall [that is much weaker in focusing medium, see Fig. \ref{fig2}(f)] may also oscillate, just like in defocusing medium. In Fig. \ref{fig3}(d) this process is accompanied by the appearance of weak modulated signal to the left of the moving domain wall. This process is exactly periodic and is not a signature of radiation from the domain wall.}

It turns out, however, that the intervals of propagation constants (defining background amplitude) over which stable topological transport is possible, are finite. In Fig.~\ref{fig3}(e) and \ref{fig3}(f) we show the examples of the unstable dynamics of the unstaggered dark soliton in defocusing medium and staggered domain wall in focusing medium. \textcolor{black}{The development of instability of the dark soliton in dynamically changing potential $\mathcal{R}$ is usually accompanied by progressively increasing rapid (on the scale of one period $Z$ of the lattice variation) oscillations of the dark notch in soliton profile. When these oscillations become sufficiently large, the soliton starts moving across the lattice with a rate substantially exceeding pumping rate and transforms into shallow gray state [see representative dynamics in Fig. \ref{fig3}(e), where rapid displacement of soliton occurs around $z\approx0.82Z$ and may not be well visible on the scale of the plot].} In the case of domain walls, \textcolor{black}{instability typically results in "jumps" of the domain wall into neighboring lattice maxima that may occur in the direction opposite to pumping direction, as shown in Fig. \ref{fig3}(f). This process is accompanied by radiation clearly visible in dynamics presented in Fig. \ref{fig3}(f) for example around $z=Z/4$. The displacement of the center of mass is not quantized in this case.}

\begin{figure}[t]
\begin{center}
			\includegraphics[width=0.5\columnwidth]{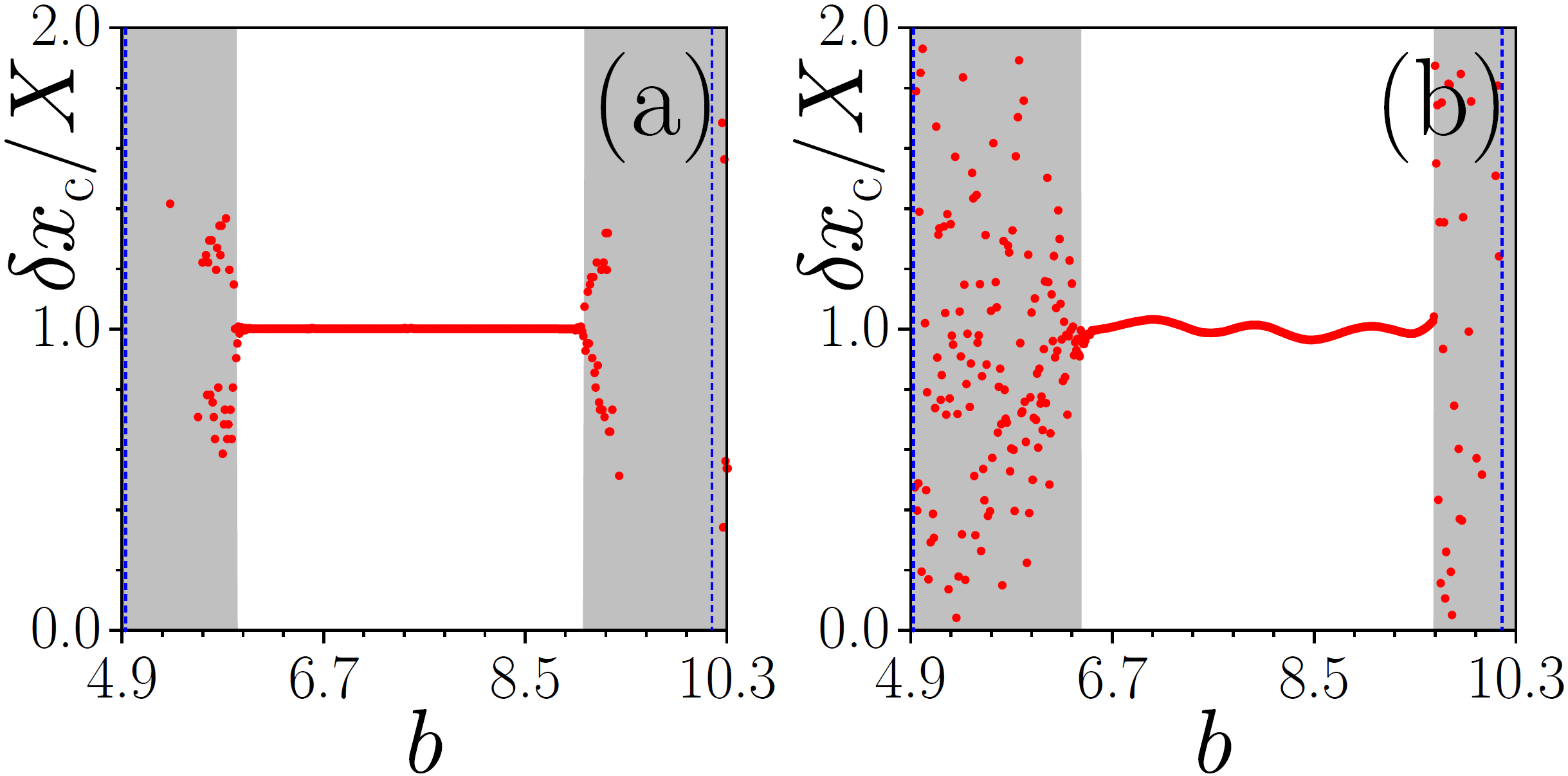}
\end{center}
\caption{Displacement of the center of solution over one pumping cycle versus propagation constant \(b\) for unstaggered dark soliton (a) and domain wall (b) at \(p_{1,2}=8\) in defocusing medium. White regions show domain of stable pumping, gray regions---of unstable pumping. Vertical blue dashed lines show borders of the first finite gap.}
\label{fig4}
\end{figure}

Typical dependence of the one-cycle spatial shift $\delta x_c$ on the propagation constant $b$ is presented in Fig.~\ref{fig4} for the unstaggered dark soliton [panel (a)] and domain wall [panel (b)] in defocusing medium. The regime of stable pumping corresponds to white domain, while in gray regions stability is lost and shift $\delta x_c$ becomes highly irregular. Notice that in defocusing medium the domain of stable pumping is located in first finite gap and it is narrower than the gap (whose borders are shown by dashed blue lines). Since the magnitude of the propagation constant is determined by the background amplitude, the existence of both lower and upper bounds for the quantized transport of dark solitons and domain walls in defocusing media is analogous to the finite admissible range of amplitudes for stable pumping of individual bright solitons in focusing media \cite{FuPRL2022a}. Notice that while instability at large values of $b$ is usually accompanied by a complete decay of soliton/domain wall, in left unstable region the propagation is accompanied by very strong radiation, especially from the regions where intensity maxima in background experience fast transverse shifts (around $z=Z/4$ and $z=3Z/4$).

\begin{figure}[t]
\begin{center}
			\includegraphics[width=0.5\columnwidth]{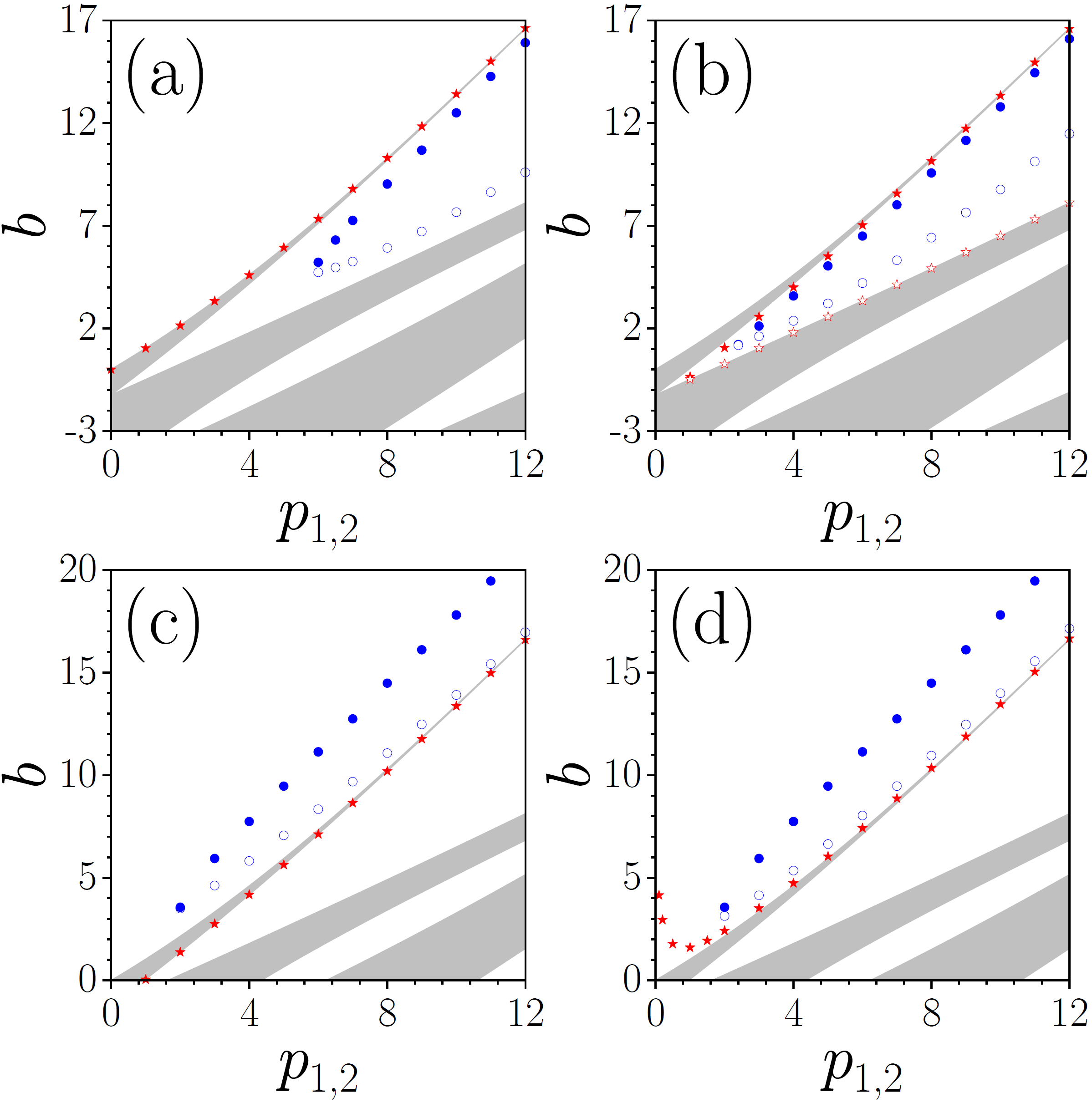}
\end{center}
\caption{Stable pumping domains on the \((p_{1,2}, b)\) plane for unstaggered dark solitons (a) and domain walls (b) in defocusing medium, and for staggered dark solitons (c) and domain walls (d) in focusing medium. Stable pumping occurs between open and solid blue dots. Gray regions show allowed bands at \(z=0\). Stars show the border of soliton existence domain. In all cases sliding velocity \(\alpha=0.001\).}
\label{fig5}
\end{figure}

More complete information on the regions of existence and stable pumping domains for different types of dark solitons and domain walls is shown in Fig.~\ref{fig5}, where the respective domains are determined as functions of the sublattice depths $p_1=p_2$. In defocusing media, where states with unstaggered background can be stable, the quantized transport of dark solitons [see Fig. \ref{fig5}(a)] and domain walls [see Fig. \ref{fig5}(b)] occurs when the propagation constants are in the first finite gap (see the open and solid blue dots showing lower and upper borders of stable pumping domain). For both dark solitons and domain walls there exists a minimal sublattice amplitude, above which stable quantized transport occurs. Interestingly, this lower limit for sublattice amplitude required for quantized transport is nearly three times smaller for domain walls than for dark solitons. At the same time, for relatively deep sublattices with $p_{1,2}\sim 12$ the domain of stable pumping is wider for dark solitons. In all cases, the region of propagation constants at which stable transport occurs, increases with sublattice amplitudes, but it always remains smaller than the existence domain of such solutions obtained for $z=0$.

Dark solitons and domain walls with staggered background in focusing medium can exhibit stable quantized transport in the semi-infinite gap [see Fig.~\ref{fig5}(c) and \ref{fig5}(d), respectively]. In this case, the width of domain of stable pumping in $b$ is similar for two different types of solutions (the upper border of stable pumping domain is actually identical, as in both cases it is determined by stability of the background, rather than by stability of the very state). However, the lower borders are different. Stable pumping is also possible starting from certain minimal value of sublattice amplitudes, that is close for dark solitons and domain walls.

\section{Conclusion}

In conclusion, we have demonstrated the principal possibility of realizing quantized transport of dark solitons and domain walls with extended background of different types. This phenomenon can occur in both defocusing and focusing optical media, where these excitations are stable when they possess unstaggered and staggered backgrounds, respectively. The quantized displacement of either a dark soliton or a domain wall over one pumping cycle is determined by the Chern number of the upper band of the linear optical potential, at least in the class of configurations considered here, in which the background is well approximated by an array of Wannier functions associated with the upper band of the instantaneous periodic potential. The interval of background amplitudes (propagation constants) where stable transport is possible is finite and broadens as the amplitude of the optical lattice is increased. Remarkably, our study shows that dark and bright solitons that arise from fundamentally different nonlinear mechanisms, exhibit in fact identical topological characteristics in the context of quantized adiabatic transport.

\section*{Acknowledgments}

Y.V.K. acknowledges funding by the research project FFUU-2024-0003 of the Institute of Spectroscopy of the Russian Academy of Sciences. 
V.V.K was supported by the Fundação para a Ciência e Tecnologia under the projects 2023.13176.PEX (DOI https://doi.org/10.54499/2023.13176.PEX) and UID/00618/2025 (DOI https://doi.org/10.54499/UID/00618/2025). F.Y. acknowledges funding by the NFSC (62575040). 



\section*{Data availability}
The data that support the findings of this study are available from the corresponding author upon request.

\section*{Code availability}
The program code used in this study is available from the corresponding authors upon request.

\end{document}